\documentclass[aps,twocolumn]{revtex4}

\usepackage{amsmath}
\usepackage{epsfig}

\begin{document}
\title{Probabilistically implementing nonlocal operation using non-maximally entangled state}
\author{Lin Chen}\email{deteriorate@zju.edu.cn}
\author{Yi-Xin Chen}\email{yxchen@zimp.edu.cn}
\affiliation{Zhejiang Insitute of Modern Physics, Zhejiang
University, Hangzhou 310027, People's Republic of China}

\begin{abstract}
We develop the probabilistic implementation of a nonlocal gate
$\exp{[i\xi{\sigma_{n_A}}\sigma_{n_B}]}$ and
$\xi\in[0,\frac\pi4]$, by using a single non-maximally entangled
state. We prove that, nonlocal gates can be implemented with a
fidelity greater than $79.3\%$ and a consumption of less than
0.969 ebits and 2 classical bits, when $\xi\leq0.353$. This
provides a higher bound for the feasible operation compared to the
former techniques \cite{Cirac,Groisman,Bennett-1}. Besides, gates
with $\xi\geq0.353$ can be implemented with the probability
$79.3\%$ and a consumption of 0.969 ebits, which is the same
efficiency as the distillation-based protocol
\cite{Groisman,Bennett-1}, while our method saves extra classical
resource. Gates with $\xi\rightarrow0$ can be implemented with
nearly unit probability and a small entanglement. We also
generalize some application to the multiple system, where we find
it is possible to implement certain nonlocal gates between many
non-entangled partners using a non-maximally multiple entangled
state.
\end{abstract}
\maketitle

Entanglement has been examined as an essential resource in most
applications of quantum information such as enhanced classical
communication, dense coding and quantum cryptography
\cite{Galindo}. Of all tasks above, the implementation of nonlocal
quantum operation on spatially distributed systems is a
considerable aspect, especially in quantum computation. This is
because, all in all, the only performance by a quantum computer is
proved to be the collective unitary operation, which is also able
to create entanglement between distributed groups. In particular,
the latter effect implies that one can realize the above tasks,
such as quantum teleportation \cite{Bennett} starting from one
entangling operation. Besides the nonlocal Hamiltonians directly
results in the dynamical evolution of distributed quantum systems,
thus it is also significative to explore other properties such as
the structure and interconvertability of nonlocal gates.

In fact, many results have been reported in this direction
\cite{Collins,Eisert,Kraus,Huelga-1,Huelga-2,Reznik,Cirac,Dur-1,Dur-2,Dur-3,Vidal-1,Groisman,Huang}.
Here, a prime problem is the efficiency for implementing a given
nonlocal gate. It is accepted that at least the consumption of 1
ebit and 2 cbits is necessarily required to faithfully implement a
general nonlocal operation \cite{Reznik}. On the other hand, a
remarkable progress has been made by \cite{Cirac}, which says
gates with small $\xi$ can be implemented by using a lower
entanglement and a more classical consumption. However the scheme
therein is inefficient for the gates with larger $\xi$, and it
usually requires an excessive amount of classical resource because
of plenty of quantum channels required. This deficiency has been
made up by the recent work in \cite{Groisman}, who employ a
probabilistic protocol to implement the nonlocal gates, with a
high fidelity when $\xi$ is small. The required classical
communication therein is 2 bits, while the entanglement can be
made arbitrarily small since it relates to the success fidelity.
However, the best attainable fidelity will lower down rapidly with
the increasing $\xi$, thus it is irresponsible to a majority of
useful gates. A higher fidelity is needed for the practical
implementation. The common weakness of above schemes is, they are
not suitable for implementing the gates with larger $\xi$.

In this paper we investigate a probabilisitic protocol to realize
the nonlocal operation
\begin{equation}
U_{AB}(\xi)=e^{i\xi\sigma_{n_A}\sigma_{n_B}}, \xi\in[0,\frac\pi4]
\end{equation}
on a general target state $\left| \Phi _{AB}\right\rangle$. We
show that, by employing the state-operator (``stator" ) approach
\cite{Reznik}, a general gate $U_{AB}$ can be implemented with the
consumption of less than 1 ebit and 2 classical bits, while the
success fidelity maintains a high level. In particular, we prove
that gates with $\xi\leq0.353$ can be implemented with a
probability greater than $79.3\%$ and a consumption of less than
0.969 ebits. The fidelity tends to one and the required
entanglement tends to zero, as $\xi$ goes to zero. This bound
effectively enlarges the region of realizable gates with a more
creditable fidelity. The above bound is universal to the residual
gates, i.e., any gate with $\xi\geq0.353$ can be implemented with
a fidelity $79.3\%$ and 0.969 ebits, which reaches the same
efficiency as the Procrustean method in \cite{Groisman,Bennett-1},
while our method is more direct and hence the extra bits are
saved. Besides, Our method realizes the same effect as that in
\cite{Groisman} when $\xi$ is small. Based on the bipartite result
above, We generalize some application to the multiple system,
where we prove that it is possible to implement certain multiple
operations between distributed $non-entangled$ systems by using a
non-maximally multiple entangled state. This effect can be
realized by adding an intermediator which we will call Charlie. If
we measure the required entanglement by dividing the multiple
systems into any bipartition, the result will be the same effect
as that in bipartite case.

Let us consider two systems A and B at different locations, the
collaborators Alice and Bob previously share an entangled state
\begin{eqnarray}
\left| \Psi
_{a{b_0}{b_1}}\right\rangle&=&\lambda_0\left|0_a0_{b_0}0_{b_1}\right\rangle+\lambda_1\left|0_a0_{b_0}1_{b_1}\right\rangle\nonumber\\
&&+\lambda_2\left|1_a1_{b_0}0_{b_1}\right\rangle+\lambda_3\left|1_a1_{b_0}1_{b_1}\right\rangle.
\end{eqnarray}
Here, particle $a$ belongs to Alice and $b_0,b_1$ belongs to Bob.
The co-efficient's $\lambda_i,i=0,1,2,3$ are non-negative and
normalized:
${\lambda^2_0}+{\lambda^2_1}+{\lambda^2_2}+{\lambda^2_3}=1$. The
entanglement is
\begin{eqnarray}
E(\left| \Psi _{a{b_0}{b_1}}\right\rangle)&=&-({\lambda^2_0}+{\lambda^2_1})\log({\lambda^2_0}+{\lambda^2_1})\nonumber\\
&&-({\lambda^2_2}+{\lambda^2_3})\log({\lambda^2_2}+{\lambda^2_3})\nonumber\\
&\equiv&-H\log{H}-(1-H)\log(1-H),
\end{eqnarray}
where $H={\lambda^2_0}+{\lambda^2_1}$, is the only parameter of
entanglement. Here we use $\log(x)$ to denote logarithms to base
2.

First we describe the general technique. Alice and Bob perform the
local unitary operation respectively
\begin{eqnarray}
U_{aA}&=&\left|0_a\right\rangle\left\langle0_a\right|\otimes{I_A}+i\left|1_a\right\rangle\left\langle1_a\right|\otimes{\sigma_{n_A}},\nonumber\\
U_{{b_1}B}&=&\left|0_{b_1}\right\rangle\left\langle0_{b_1}\right|\otimes{I_B}+\left|1_{b_1}\right\rangle\left\langle1_{b_1}\right|\otimes{\sigma_{n_B}}.
\end{eqnarray}
Then Alice measures ${\sigma_x}_a$ and transmits the result to Bob
by sending 1 classical bit. Following this message Bob will do
nothing or ${\sigma_z}_{b_0}$. Thus they get an initial stator
\begin{eqnarray}
S_{ini}&=&\lambda_0\left|0_{b_0}0_{b_1}\right\rangle{I_A}{I_B}+\lambda_1\left|0_{b_0}1_{b_1}\right\rangle{I_A}{\sigma_{n_B}}\nonumber\\
&&+i\lambda_2\left|1_{b_0}0_{b_1}\right\rangle{\sigma_{n_A}}{I_B}+i\lambda_3\left|1_{b_0}1_{b_1}\right\rangle{\sigma_{n_A}}{\sigma_{n_B}}.
\end{eqnarray}
Now Bob collectively measures particle $b_0$ and $b_1$ in the
following basis
\begin{eqnarray}
\left|B_{00}\right\rangle&=&\cos{\delta_0}\left|00\right\rangle+\sin{\delta_0}\left|11\right\rangle,\nonumber\\
\left|B_{01}\right\rangle&=&\cos{\delta_0}\left|11\right\rangle-\sin{\delta_0}\left|00\right\rangle,\nonumber\\
\left|B_{10}\right\rangle&=&\cos{\delta_1}\left|01\right\rangle+\sin{\delta_1}\left|10\right\rangle,\nonumber\\
\left|B_{11}\right\rangle&=&\cos{\delta_1}\left|10\right\rangle-\sin{\delta_1}\left|01\right\rangle.
\end{eqnarray}
The corresponding probability to get each Basis is
\begin{eqnarray}
P(\left|B_{00}\right\rangle)&=&{\lambda^2_0}{\cos^2{\delta_0}}+{\lambda^2_3}{\sin^2{\delta_0}},\nonumber\\
P(\left|B_{01}\right\rangle)&=&{\lambda^2_0}{\sin^2{\delta_0}}+{\lambda^2_3}{\cos^2{\delta_0}},\nonumber\\
P(\left|B_{10}\right\rangle)&=&{\lambda^2_1}{\cos^2{\delta_1}}+{\lambda^2_2}{\sin^2{\delta_1}},\nonumber\\
P(\left|B_{11}\right\rangle)&=&{\lambda^2_1}{\sin^2{\delta_1}}+{\lambda^2_2}{\cos^2{\delta_1}}.
\end{eqnarray}
After getting one of the resulting operators, Bob broadcasts 1 bit
to inform Alice that she will perform $\sigma_{n_A}$ or nothing.
With the local operation $\sigma_{n_B}$ or $I_B$ by Bob himself,
the operators they obtain will be respectively
\begin{eqnarray}
S(\left|B_{00}\right\rangle)&=&{\lambda_0}{\cos{\delta_0}}{I_A}{I_B}+i{\lambda_3}{\sin{\delta_0}}{\sigma_{n_A}}{\sigma_{n_B}},\nonumber\\
S(\left|B_{01}\right\rangle)&=&{\lambda_3}{\cos{\delta_0}}{I_A}{I_B}+i{\lambda_0}{\sin{\delta_0}}{\sigma_{n_A}}{\sigma_{n_B}},\nonumber\\
S(\left|B_{10}\right\rangle)&=&{\lambda_1}{\cos{\delta_1}}{I_A}{I_B}+i{\lambda_2}{\sin{\delta_1}}{\sigma_{n_A}}{\sigma_{n_B}},\nonumber\\
S(\left|B_{11}\right\rangle)&=&{\lambda_2}{\cos{\delta_1}}{I_A}{I_B}+i{\lambda_1}{\sin{\delta_1}}{\sigma_{n_A}}{\sigma_{n_B}}.
\end{eqnarray}
We show the following results based on the above argument.

(result 1) A simple observation is that, if we set
$\lambda_0=\lambda_3$ and $\lambda_1=\lambda_2$, while
$\delta_0=\delta_1=\xi$, which can be decided by Bob. Thus we have
faithfully realized the target operation
$U_{AB}(\xi)=e^{i\xi\sigma_{n_A}\sigma_{n_B}}$. However we readily
get another result, i.e., $H=\frac12$, which means that the
required entanglement is maximal. Since the classical consumption
in this process is 2 bits, we thus reach the same efficiency in
\cite{Reznik}.

(result 2) On the other hand, if we discard half of the resulting
operators above, e.g., we only require $\lambda_0=\lambda_3$ and
$\delta_0=\xi$. As described above, this fidelity is
\begin{equation}
P(\left|B_{00}\right\rangle)+P(\left|B_{01}\right\rangle)=2{\lambda^2_0}.
\end{equation}
if we set $\lambda_1=0$, thus we could get the same efficiency as
that of Procrustean method in \cite{Bennett-1}, where a maximally
entangled state will be generated in the intermediate stage with a
fidelity, and this EPR singlet is used to implement a general
operation \cite{Groisman}. Therefore we can implement the nonlocal
gate $U_{AB}(\xi)$ with a fidelity $2{\lambda^2_0}$ and the
entanglement required is $-{\lambda^2_0}\log{\lambda^2_0}
-(1-{\lambda^2_0})\log(1-{\lambda^2_0})$. Besides, our method is
more direct than the Procrustean method and thus economizes the
extra classical consumption. We call this technique
Fast-Procrustean-Transformation (FPT).

(result 3) We adopt another path to explain how to implement gates
with nearly unit probability and a small entanglement, when
$\xi\rightarrow0$. Here we set $\lambda_0=\lambda_3=0$ and
$\lambda_1=\cos\alpha$, $\lambda_2=\sin\alpha$. Following the
above procedure we will get one of the two operators
$S(\left|B_{10}\right\rangle)$ and $S(\left|B_{11}\right\rangle)$.
The probability to get $S(\left|B_{10}\right\rangle)$ is
\begin{equation}
P(\left|B_{10}\right\rangle)={\cos^2\alpha}
{\cos^2{\delta_1}}+{\sin^2\alpha}{\sin^2{\delta_1}}.
\end{equation}
In order to realize $U_{AB}(\xi)$, we set
\begin{equation}
\frac{{\sin\alpha}\sin{\delta_1}}{{\cos\alpha}\cos{\delta_1}}=\tan\xi.
\end{equation}
Using this condition we get
\begin{eqnarray}
P(\left|B_{10}\right\rangle)&=&\frac{{\sec^2\xi}}{{\sec^2\alpha}+{\csc^2\alpha}{{\tan^2\xi}}}\nonumber\\
&\leq&\frac{1}{1+\sin{2\xi}},
\end{eqnarray}
where the equality holds when $\tan\alpha=\sqrt{\tan\xi}$. The
entanglement resource required for this optimal fidelity is
\begin{eqnarray}
E(\left| \Psi
_{a{b_0}{b_1}}\right\rangle)&=&-\frac{1}{1+\tan\xi}\log(\frac{1}{1+\tan\xi})\nonumber\\
&&-\frac{\tan\xi}{1+\tan\xi}\log(\frac{\tan\xi}{1+\tan\xi}).
\end{eqnarray}
Therefore, we can implement the gate $U_{AB}(\xi)$ with a high
fidelity and very small entanglement, when $\xi$ is near zero.
This process achieves the same effect as that of \cite{Groisman}
and thus it is more efficient than that of \cite{Cirac} for the
much saving classical consumption.

In the following, we analyze the implementation of
$U_{AB}(\xi)=e^{i\xi\sigma_{n_A}\sigma_{n_B}}$, when $\xi$ is a
general amount. We show that, combined with the FPT technique, it
is feasible to implement a general nonlocal gate with a high
fidelity, by using less than 1 ebit and 2 bit of classical
communication.

We still start from the state $ \left| \Psi
_{a{b_0}{b_1}}\right\rangle=\lambda_0\left|0_a0_{b_0}0_{b_1}\right\rangle+\lambda_1\left|0_a0_{b_0}1_{b_1}\right\rangle
+\lambda_2\left|1_a1_{b_0}0_{b_1}\right\rangle+\lambda_3\left|1_a1_{b_0}1_{b_1}\right\rangle$.
Here, the co-efficient's are defined as
\begin{eqnarray}
\lambda_0&=&\frac{\tan{\theta_1}\cos{\theta_1}}{\sqrt{(\tan^2{\theta_0}+\tan^2{\theta_1})(\cos^2{\theta_0}+\cos^2{\theta_1})}},\nonumber\\
\lambda_1&=&\frac{\tan{\theta_0}\cos{\theta_1}}{\sqrt{(\tan^2{\theta_0}+\tan^2{\theta_1})(\cos^2{\theta_0}+\cos^2{\theta_1})}},\nonumber\\
\lambda_2&=&\frac{\tan{\theta_0}\cos{\theta_0}}{\sqrt{(\tan^2{\theta_0}+\tan^2{\theta_1})(\cos^2{\theta_0}+\cos^2{\theta_1})}},\nonumber\\
\lambda_3&=&\frac{\tan{\theta_1}\cos{\theta_0}}{\sqrt{(\tan^2{\theta_0}+\tan^2{\theta_1})(\cos^2{\theta_0}+\cos^2{\theta_1})}},
\end{eqnarray}
where the parameters $\theta_0,\theta_1\in[0,\frac\pi2]$, and
$H=\frac{\cos^2{\theta_1}}{\cos^2{\theta_0}+\cos^2{\theta_1}}$.
First Bob implements a POVM operation on particle $b_0$ as follows
\[
M_0=\left(\begin{array}{cc} \cos{\theta_0} & 0\\
0 & \cos{\theta_1}\end{array}\right),
M_1=\left(\begin{array}{cc} \sin{\theta_0} & 0\\
0 & \sin{\theta_1}\end{array}\right).
\]
We notice that
$\lambda_0\cos{\theta_0}=\lambda_3\cos{\theta_1}$,$\lambda_1\cos{\theta_0}=\lambda_2\cos{\theta_1}$.
Thus if Bob implements $M_0$ and we restart from state
${M_0}_{b_0}\left| \Psi _{a{b_0}{b_1}}\right\rangle$. Following
the general technique above, one readily finds this is the case in
result (1), i.e., $\lambda_0\cos{\theta_0}\rightarrow\lambda_0$,
$\lambda_3\cos{\theta_1}\rightarrow\lambda_3$, etc, which means we
have faithfully implemented the gate $U_{AB}(\xi)$. On the other
hand if Bob implements $M_1$, then he performs a CNOT gate on
particle $b_0$ and $b_1$. Thus the resulting state is
\begin{eqnarray*}
\left|\Psi^{res}_{a{b_0}{b_1}}\right\rangle&=&\lambda_0\sin{\theta_0}\left|0_a0_{b_0}0_{b_1}\right\rangle+\lambda_1\sin{\theta_0}\left|0_a0_{b_0}1_{b_1}\right\rangle\\
&&+\lambda_3\sin{\theta_1}\left|1_a1_{b_0}0_{b_1}\right\rangle+\lambda_2\sin{\theta_1}\left|1_a1_{b_0}1_{b_1}\right\rangle.
\end{eqnarray*}
Again Alice and Bob follow the general technique, starting with
the state $\left|\Psi^{res}_{a{b_0}{b_1}}\right\rangle$. Notice
that $\lambda_0\sin{\theta_0}=\lambda_2\sin{\theta_1}$ and Bob
chooses $\delta_0=\xi$, thus we realize $U_{AB}(\xi)$ on operators
$S(\left|B_{00}\right\rangle)$ and $S(\left|B_{01}\right\rangle)$.
Furthermore, we can make operator
$S(\left|B_{10}\right\rangle)=U_{AB}(\xi)$ by supposing
\begin{equation}
\frac{\lambda_3\sin{\theta_1}\sin{\delta_1}}{\lambda_1\sin{\theta_0}\cos{\delta_1}}=\tan\xi,
\end{equation}
or
\begin{equation}
n\tan{\xi}=\tan{\delta_1},
n\equiv\frac{\tan^2{\theta_0}}{\tan^2{\theta_1}}.
\end{equation}
Therefore the only failure case is $S(\left|B_{11}\right\rangle)$,
whose probability is
\begin{eqnarray}
P(\left|B_{11}\right\rangle)&=&\lambda^2_1\sin^2{\theta_0}\sin^2{\delta_1}+\lambda^2_3\sin^2{\theta_1}\cos^2{\delta_1}\nonumber\\
&\equiv&\frac{1+n^4{\tan^2{\xi}}}{(1+n^2{\tan^2{\xi}})(1+n)(1+nb)}.
\end{eqnarray}
Here $b\equiv2\cot^2{\theta_0}+1$, which should be near one by the
fact that
\begin{equation}
H=\frac{\cos^2{\theta_1}}{\cos^2{\theta_0}+\cos^2{\theta_1}}=(\frac{1-n^{-1}}{1+\frac{2}{b-1}}+1+n^{-1})^{-1}.
\end{equation}
That is, the minimum value of $E(\left| \Psi
_{a{b_0}{b_1}}\right\rangle)$ will be attained as $b$ tends to
one. In what follows we will keep to this principle in order to
make a lowest consumption of entanglement. We take $n$ as the
independent variable and calculate the minimum value of
$P(\left|B_{11}\right\rangle)$ as follows
\begin{equation*}
\frac{d}{dn}P(\left|B_{11}\right\rangle)=\frac{C_0\tan^4{\xi}+C_1\tan^2{\xi}+C_2}{(1+n^2{\tan^2{\xi}})^2(1+n)^2(1+nb)^2},
\end{equation*}
where
\begin{eqnarray}
C_0&=&(2+n+nb)n^5,\nonumber\\
C_1&=&(2n^3+3n^2-4n-3)n^2b+(3n^3+4n^2-3n-2)n,\nonumber\\
C_2&=&-1-b-2nb.
\end{eqnarray}
We now explore some conclusions. First, it is easy to check that
we have to provide more entanglement than the FPT technique both
under an identical success probability when $n\in[0,n_0],
n_0\simeq1.214$, which satisfies $n^6+2n^5+3n^4-4n^3-3n^2-2n-1=0$.
One can compare the two amounts of $E(\left| \Psi
_{a{b_0}{b_1}}\right\rangle)$ with
$H_0=\frac{1-P(\left|B_{11}\right\rangle)}{2}$ and
$H_1=(1+n^{-1})^{-1}$ to clarify this result. Second, one can
readily checks that $C_2<0$ and $C_0+C_1+C_2>0$ when $n\geq{n_0}$.
Under this condition there exists certain $\xi$ to satisfy
$C_0\tan^4{\xi}+C_1\tan^2{\xi}+C_2=0$, and this is indeed the
minimum value of $P(\left|B_{11}\right\rangle)$. Again we need to
compare this improved technique with the FPT technique under an
identical fidelity, which we describe in Figure 1. The parameters
therein are as follows (We set $b=1.001$ in this figure and in
fact this parameter tends to one. Thus we take the following
expression of $E_0$, which approximates the image in figure 1).
\begin{eqnarray}
E_{FPT}&=&-{H_0}\log{H_0}-(1-H_0)\log(1-H_0),\nonumber\\
E_0&=&-{H_1}\log{H_1}-(1-H_1)\log(1-H_1),\nonumber\\
\tan^2{\xi}&=&-\frac{C_1}{2C_0}+\sqrt{\frac{C^2_1}{4C^2_0}-\frac{C_2}{C_0}},\nonumber\\
F&=&1-P(\left|B_{11}\right\rangle).
\end{eqnarray}
From the comparison we see that it is more efficient to employ the
improved technique to implement the gate $U_{AB}(\xi)$ when
$\xi\leq0.353$. On the other hand, one still employs the FPT
technique when $\xi>0.353$. Since the FPT technique is universal
to any $\xi$, we thus get the conclusion that any gate
$U_{AB}(\xi)$ with $\xi>0.353$ can be implemented by the
consumption of 0.969 ebit and 2 cbits, under a fidelity $79.3\%$.
One will see the success fidelity we obtain is much higher than
that in \cite{Groisman}, while our entanglement consumption is
also lower than one ebit and tends to zero as $\xi$ goes to zero.
Furthermore, compared to \cite{Cirac} our method saves
entanglement when $\xi$ increases, e.g., when $\xi=0.17$ the
required ebit here is 0.897 with a fidelity $85.6\%$, while it
will be 1.016 ebits in \cite{Cirac}.

\begin{figure}[ht]
\epsfxsize=7.8cm \epsfbox{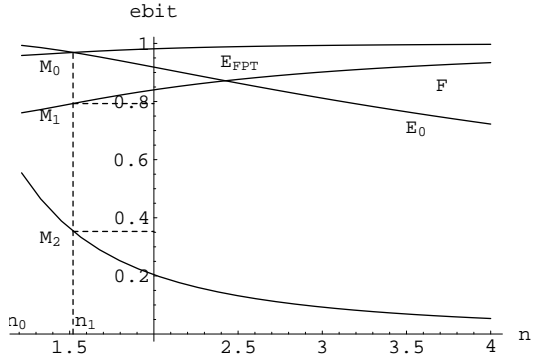} \caption{Comparison of
entanglement consumption $E_0$ and $E_{FPT}$ between the improved
technique and the FPT technique, under a common success fidelity
$F=1-P(\left|B_{11}\right\rangle)$ and $b=1.001$. The lowest curve
represents $\xi$, which increases as $n$ tends to $n_0$.
$E_{FPT}=E_0\simeq0.969$ at the point $M_0$, where
$n_1\simeq1.521$. The lowest fidelity $F=0.793$, is obtained at
the point $M_1$. The corresponding value of $\xi$ approximates
0.353 at the point $M_2$.}
\end{figure}

In the following we generalize some useful application to the
multiple gate. First we review the general technique, where we
employ the entangled state $\left| \Psi
_{a{b_0}{b_1}}\right\rangle=\lambda_0\left|0_a0_{b_0}0_{b_1}\right\rangle+\lambda_1\left|0_a0_{b_0}1_{b_1}\right\rangle
+\lambda_2\left|1_a1_{b_0}0_{b_1}\right\rangle+\lambda_3\left|1_a1_{b_0}1_{b_1}\right\rangle$.
Now we introduce an assistant called Charlie and the three
partners share such entangled state instead
\begin{eqnarray}
\left| \Psi_{ab{c_0}{c_1}}\right\rangle&=&\lambda_0\left|0_a0_b0_{c_0}0_{c_1}\right\rangle+\lambda_1\left|0_a0_b0_{c_0}1_{c_1}\right\rangle\nonumber\\
&&+\lambda_2\left|1_a1_b1_{c_0}0_{c_1}\right\rangle+\lambda_3\left|1_a1_b1_{c_0}1_{c_1}\right\rangle,
\end{eqnarray}
where particle $a$ belongs to Alice, $b$ belongs to Bob and
Charlie possesses $c_0$ and $c_1$. The difference between $\left|
\Psi_{ab{c_0}{c_1}}\right\rangle$ and $\left| \Psi
_{a{b_0}{b_1}}\right\rangle$ is: in state $\left|
\Psi_{ab{c_0}{c_1}}\right\rangle$, the assistant Charlie replaces
the role of Bob in state $\left| \Psi
_{a{b_0}{b_1}}\right\rangle$. Therefore we can carry out a scheme
similar to that of general technique by using $\left|
\Psi_{ab{c_0}{c_1}}\right\rangle$ like this. First Alice and Bob
locally perform $U_{aA}$ and $U_{bB}$ on their particles
respectively, followed by the measurement of ${\sigma_x}_a$ and
${\sigma_x}_b$. Now they get another initial stator
\begin{eqnarray}
S^{\prime}_{ini}&=&\lambda_0\left|0_{c_0}0_{c_1}\right\rangle{I_A}{I_B}+\lambda_1\left|0_{c_0}1_{c_1}\right\rangle{I_A}{I_B}\\
&&+i\lambda_2\left|1_{c_0}0_{c_1}\right\rangle{\sigma_{n_A}}\sigma_{n_B}+i\lambda_3\left|1_{c_0}1_{c_1}\right\rangle{\sigma_{n_A}}{\sigma_{n_B}}.\nonumber
\end{eqnarray}
The remainder of this process is the collective measurement on
particle $c_0$ and $c_1$ by Charlie and all analysis above works.
In the case of the improved technique, the POVM and CNOT gate
therein will be implemented by Charlie instead. This completes the
whole argument using state $\left|
\Psi_{ab{c_0}{c_1}}\right\rangle$. Therefore it is also feasible
to employ the state $\left| \Psi_{ab{c_0}{c_1}}\right\rangle$ to
implement the nonlocal gate
$U_{AB}(\xi)=e^{i\xi\sigma_{n_A}\sigma_{n_B}},
\xi\in[0,\frac\pi4]$, while the classical consumption is 4 bits
here. However one will notice that, there is no entanglement
between any two partners in the state $\left|
\Psi_{ab{c_0}{c_1}}\right\rangle$, in fact any two of them are
separable \cite{Peres}. This is similar to the GHZ qutrit and thus
we call $\left| \Psi_{ab{c_0}{c_1}}\right\rangle$ the $quasi-GHZ$
state. As we know, the GHZ state is the maximally entangled state
in multiple system. Here we regard the $quasi-GHZ$ state as the
$non-maximally$ entangled state in multiple system by diving the
multiple system into a random bipartite system, namely A-BC, B-AC
or C-AB. One readily gets the entropy of any system above will be
$-H\log{H}-(1-H)\log(1-H)$, which is less than one except
$H=\frac12$. Since $H\neq\frac12$ in the bipartite results, we
also employ the non-maximally entangled state in the multiple case
\cite{notation1}.

The above argument implies that one can implement the nonlocal
gate on a pair of partners who are not entangled by adding a third
company Charlie, who indeed works as an $intermediator$. The major
virtue of this mode is, Alice and Bob will only need to know the
content of ${\sigma_n}_A$ and ${\sigma_{n^\prime}}_B$
respectively(we notice it is not necessarily that the two
directions $n$'s are always the same). Therefore, the knowledge of
$\xi$ can be unclear to both Alice and Bob, while Bob must hold
all knowledge of the state $\left| \Psi
_{a{b_0}{b_1}}\right\rangle$ and $\xi$, ${\sigma_n}_B$ in the
general technique above, which is also a universal status in
existed work. Second, if Charlie performs a local operation
\begin{equation}
U_{{c_0}C}=\left|0_{c_0}\right\rangle\left\langle0_{c_0}\right|\otimes{I_C}+\left|1_{c_0}\right\rangle\left\langle1_{c_0}\right|\otimes{\sigma_{n_C}}
\end{equation}
on the stator $S^{\prime}_{ini}$ where $C$ is the target particle
of Charlie, this will lead to the realization of gate
$U_{ABC}(\xi)=e^{i\xi\sigma_{n_A}\sigma_{n_B}\sigma_{n_C}},
\xi\in[0,\frac\pi4]$. One can also generalize the $quasi-GHZ$
state to a multiple case, that is
\begin{eqnarray}
\left| \Psi_{a_1a_2...a_N{c_0}{c_1}}\right\rangle&=&\lambda_0\left|0_{a_1}0_{a_2}...0_{a_N}0_{c_0}0_{c_1}\right\rangle\nonumber\\
&&+\lambda_1\left|0_{a_1}0_{a_2}...0_{a_N}0_{c_0}1_{c_1}\right\rangle\nonumber\\
&&+\lambda_2\left|1_{a_1}1_{a_2}...1_{a_N}1_{c_0}0_{c_1}\right\rangle\nonumber\\
&&+\lambda_3\left|1_{a_1}1_{a_2}...1_{a_N}1_{c_0}1_{c_1}\right\rangle.
\end{eqnarray}
Here N particles $a_i,i=1,2,...,N$ are spatially distributed in N
partners respectively, which we call $A_i,i=1,2,...,N$. Following
the same technique above, we are able to realize the gate
$U_{A_1A_2...A_NC}(\xi)=\exp[{i\xi\sigma_{n_{A_1}}\sigma_{n_{A_2}}...\sigma_{n_{A_N}}\sigma_{n_C}}],
\xi\in[0,\frac\pi4]$, where the classical consumption will be $2N$
bits which equals to the amount in \cite{Reznik}. Again, each
participant only needs to know the corresponding gate $\sigma_n$,
except Charlie. Besides, comparing with the technique in
\cite{Reznik} our scheme reduces the entanglement consumption for
implementing the gate $U_{A_1A_2...A_NC}(\xi)$, if we divide the
multiple system into random bipartite system. In particular, the
entanglement between Charlie and the N distributed receivers
remains less than 1 ebit in our scheme, while it requires N ebits
in \cite{Reznik}.

In conclusion, we have given an explicit scheme to realize the
nonlocal operation $U_{AB}(\xi)=e^{i\xi\sigma_{n_A}\sigma_{n_B}},
\xi\in[0,\frac\pi4]$ by using a single non-maximally entangled
state. The technique introduced here allows one to implement the
nonlocal gate with the consumption of less than 0.969 ebits and 2
cbits, and the minimum success fidelity is $79.3\%$. The
entanglement consumption will decrease to zero and the fidelity
goes to one as $\xi$ ranges from 0.353 to zero, which enlarges the
region of applicable gates and effectively saves entanglement and
operates more reliably compared to the former techniques. The
classical communication is also the lowest amount so far. An open
question is, whether there is more efficient technique than the
FPT way in the region $\xi\geq0.353$. The required entanglement
here can be further reduced if lower fidelity is allowed. We also
provide another method to perfectly implement the gate with a
small $\xi$, which reaches the same effect in \cite{Groisman}.
Besides, we generalize this technique to the implementation of
multiple gate also using non-maximally multiple entangled state
and in a sense it largely reduces the required entanglement in the
multiple case. The fact that nonlocal gate could be performed
between non-entangled partners is also notable, which could be
useful to other tasks such as quantum secret sharing.

The work was partly supported by the NNSF of China (Grant
No.90203003), NSF of Zhejiang Province (Grant No.602018), and by
the Foundation of Education Ministry of China (Grant
No.010335025).


\begin{references}
\bibitem{notation1}It is
noticeable that, so far there is no specific definition of
non-maximally entangled state in multiple system. In this paper
the employed entangled state has a general form $\left|
\Psi_{a_1a_2...a_N{c_0}{c_1}}\right\rangle$ described in the text,
which could be made out from a quNit
$\sqrt{{\lambda^2_0}+{\lambda^2_1}}\left|0_{a_1}0_{a_2}...0_{a_N}0_{c_0}0_{c_1}\right\rangle
+\sqrt{{\lambda^2_2}+{\lambda^2_3}}\left|1_{a_1}1_{a_2}...1_{a_N}1_{c_0}1_{c_1}\right\rangle$,
by locally collective unitary operation by Charlie. This process
preserves the entanglement. It is easy to see that this quNit is
equivalent to the standard N-GHZ state under SLOCC
\cite{Bennett-2,Dur-4}. Therefore it is reasonable to compare the
entanglement between the above quNit and the N-GHZ state, by
dividing the multiple system into any bipartite system including
the same parties. As described in the text(the example of N=2),
this quNit is less entangled than the GHZ state and it is
non-maximally.
\end{references}
\end{document}